\documentclass[11pt,twoside]{article}
\usepackage{asp2010}

\resetcounters

\begin{document}

\title{Host Galaxy Morphology and the AGN Unified Model}

\author{Jonathan R. Trump
\affil{University of California Observatories/Lick Observatory and
  Department of Astronomy and Astrophysics, University of California,
  Santa Cruz, CA 95064}}

\begin{abstract}
  We use a sample of active galaxies from the Cosmic Evolution Survey
  to show that host galaxy morphology is tied to the accretion rate
  and X-ray obscuration of its active galactic nucleus (AGN).
  Unobscured and rapidly accreting broad-line AGNs are more likely to
  be in spheroid-dominated hosts than weak or obscured AGNs, and
  obscured AGNs are more likely to have disturbed host galaxies.  Much
  of the disagreement in previous work on the AGN-merger connection is
  likely due to each study probing AGNs with different obscuration and
  accretion properties.  Instead, only obscured AGNs seem to
  merger-driven, while weak AGNs are fed by stochastic processes in
  disks and rapidly-accreting broad-line AGNs require massive bulges.
  Our observed ``unified model'' for AGN hosts fits with theoretical
  models for merger-driven AGN evolution, but is also consistent with
  steady-state AGN activity.
\end{abstract}

\section{Introduction}

The most popular theoretical framework for coevolving active galactic
nuclei (AGNs) and galaxies invokes major mergers to fuel both
starbursts and quasars \citep{san88,hop06}.  But there also exist
secular processes which can grow AGNs in disks, through stochastic gas
accretion and stellar mass loss \citep{hh06} or cold streams and disk
instabilities \citep{bou11}.  The relative dominance of these fueling
modes may evolve: perhaps secular processes are efficient in the
gas-rich $z>0.5$ universe but mergers are required to funnel gas to
galaxy nuclei at $z \sim 0$, in a way analogous to star-forming
galaxies \citep[e.g.][]{gen10}.

\begin{figure}[t]
\begin{center}
 \scalebox{0.8}
  {\plotone{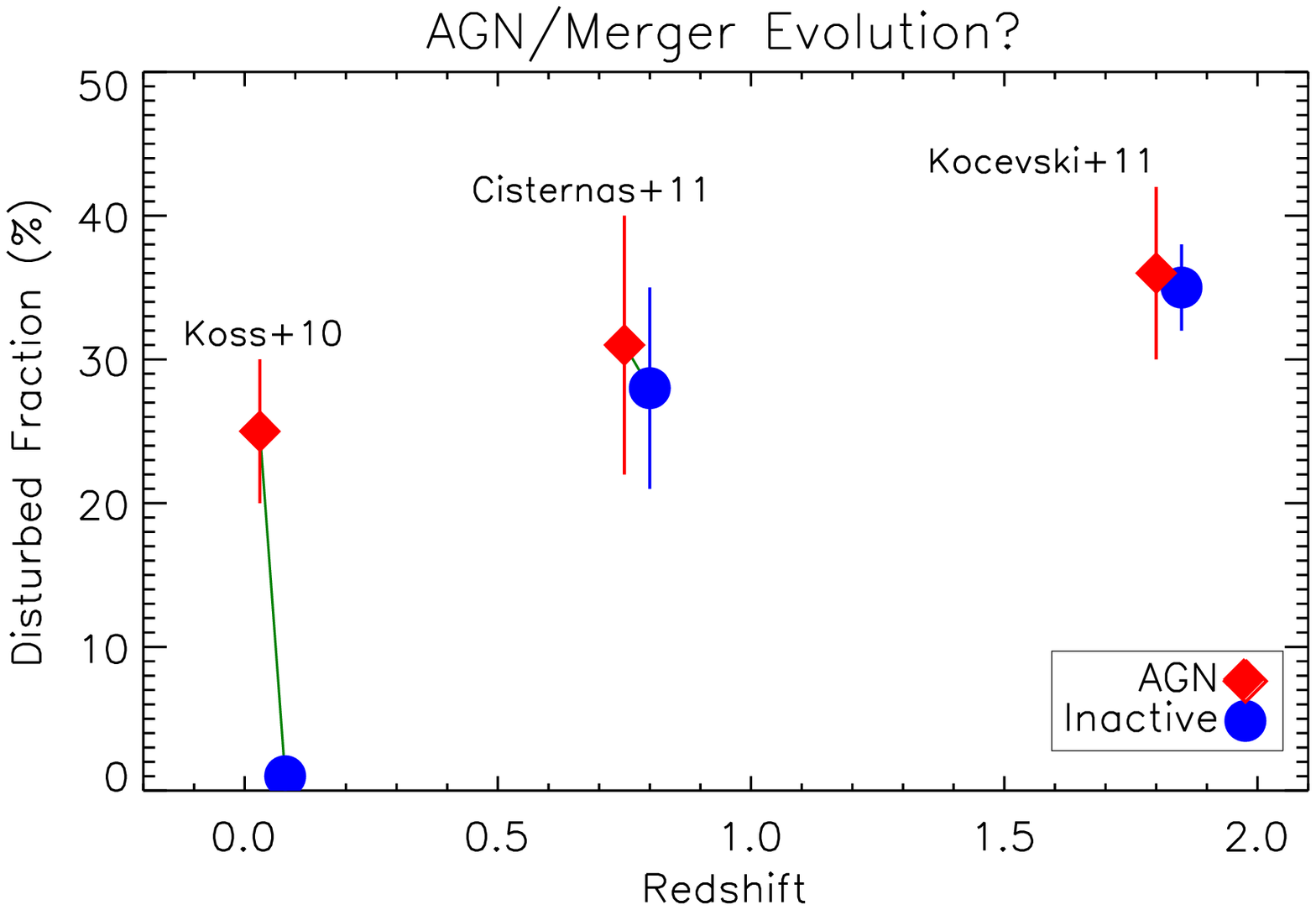}} 
\end{center}
\caption{The fraction of AGNs and inactive galaxies which are
  disturbed, measured from X-ray selected samples at three different
  redshifts: $z \sim 0.03$ \citep{koss10}, $z \sim 0.75$
  \citep{cis11}, and $z \sim 1.8$ \citep{koc11}.  At first glance, it
  appears that the AGN-merger connection evolves, with mergers driving
  AGNs locally but not in the past.  However the apparent trend is
  probably caused by differences in the AGN samples rather than true
  evolution.
\label{fig:mergerfrac}}
\end{figure}

Observations of AGN host galaxy morphologies have the potential to
distinguish secular processes from merger fueling.  Figure
\ref{fig:mergerfrac} shows examples of host morphology studies from
X-ray AGNs at three different redshifts: $z \sim 0.03$ \citep{koss10},
$z \sim 0.75$ \citep{cis11}, and $z \sim 1.8$ \citep{koc11}.  In each
survey, AGNs and mass-matched inactive galaxies are blindly classified
by visual inspection.  Naively comparing the three results suggests
evolution in the importance of merger fueling for AGNs, but each study
is biased to selecting different kinds of AGNs.

Contrary to the classical unified model \citep{ant93}, recent work has
shown that rapidly accreting quasars have very different fueling and
feedback modes from weak AGNs \citep{ho08,tru11,ant11}.  In
particular, \citet{tru11} shows that broad-line quasars have strong
radiative winds and luminous accretion disks, while weakly accreting
narrow-line AGNs have radiatively inefficient accretion flows and
radio-mode feedback.  Understanding AGN-galaxy coevolution then
requires studying host galaxy type across the two axes of AGN
accretion rate and obscuration.  Yet previous AGN host studies
(including those shown in Figure \ref{fig:mergerfrac}) have generally
not distinguished between weakly accreting LINERs, Compton-thick AGNs,
and powerful quasars.

Here we study host galaxy morphologies for 70 X-ray selected AGNs from
the Cosmic Evolution Survey \citep[COSMOS,][]{sco07}.  These AGNs span
three orders of magnitude in each of accretion rate and column
density, and their observed data demonstrate that host galaxy type is
connected to AGN properties.

\section{Observational Data}

We study AGN and host properties for the 70 COSMOS sources which have:
\begin{enumerate}
  \item XMM point source detections from \citet{bru10}, with $L_X>3
    \times 10^{42}$~erg~s$^{-1}$ and $>40$ X-ray counts for accurate
    $N_H$ estimates.
  \item High-confidence ($>90\%$ likelihood as correct) redshifts and
    classification from optical spectroscopy \citep{tru09a}.
  \item Rest-frame optical host galaxy morphologies from
    \citet{gab09}, restricted to $z<1$ AGNs with {\it HST}/ACS imaging
    \citep{koe07}.
\end{enumerate}
The 70 AGNs include 5 broad-line, 38 narrow-line, and 27 optically
dull AGNs.

We parameterize AGN accretion rate using the Eddington ratio,
$\lambda_{Edd} \equiv L_{bol}/L_{Edd} \sim L_{bol}/M_{BH}$.
\citet{tru11} already estimated $\lambda_{Edd}$ for unobscured
($N_H<10^{22}$~cm$^{-2}$) AGNs in our sample, with $L_{bol}$ from
model fits to the multiwavelength SEDs and $M_{BH}$ from broad
emission lines (when available) or the $M_{BH}-M_*$ relation.  For
obscured ($N_H>10^{22}$~cm$^{-2}$) AGNs, we cannot estimate $L_{bol}$
in the same fashion because most of the SED is heavily extincted.
Instead we estimate $L_{bol}$ using an infrared bolometric correction,
$L_{bol}=8L_{6{\mu}m}$ \citep{ric06}.  Note that for unobscured AGNs,
this bolometric correction is roughly consistent with $L_{bol}$
measured from the SED method.  For all AGNs $N_H$ is estimated from
the X-ray spectra, using either the XMM-COSMOS data or the deeper {\it
  Chandra} data \citep{elv09} when available.

Morphology measurements for the host galaxies of the 70 AGNs are
presented by \citet{gab09}.  Briefly, {\it HST}/ACS images were fit
with GALFIT \citep{peng02}, using a point source component to describe
the AGN and a single \citet{ser68} function to describe the extended
galaxy.  For many of narrow-line and lineless AGNs, a better fit was
obtained without the point source component: for these systems we use
the S\'{e}rsic-only fit of \citet{gab09}.  In addition to the
S\'{e}rsic index $n$, \citet{gab09} also measure the asymmetry
parameter $A$.  We use $n$ and $A$ to group AGN host galaxies into
three categories: $A>0.5$ galaxies are disturbed, $n>2.5$ galaxies are
spheroid-dominated, and $n<2.5$ galaxies are disks.

\section{The Host Galaxies of Active Galactic Nuclei}

\begin{figure}[t]
\begin{center}
 \scalebox{1.0}
  {\plotone{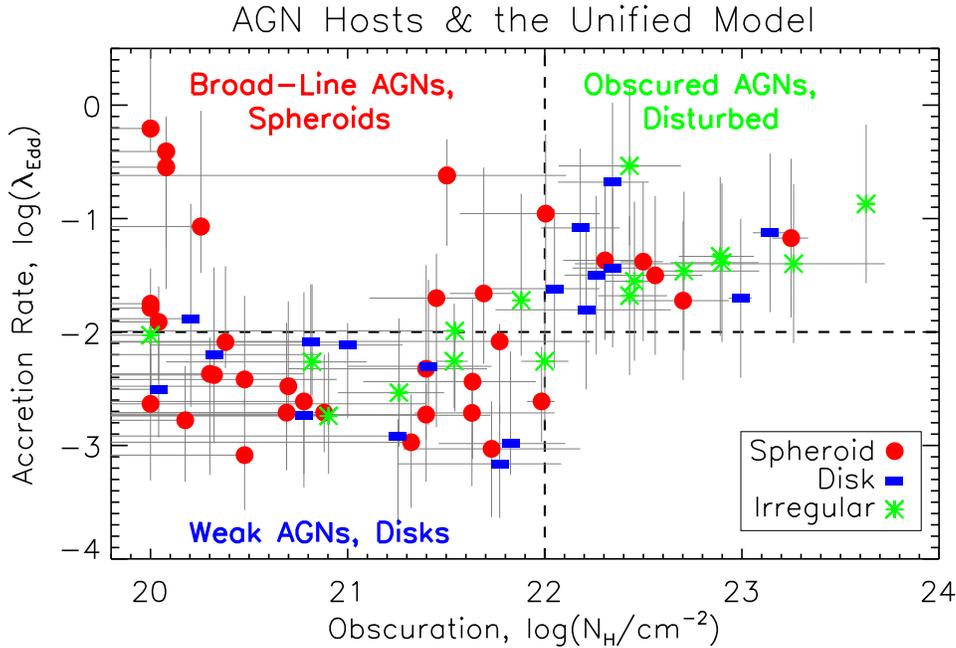}}
\end{center}
\caption{Accretion rate and obscuration for the AGNs in our sample,
  with host galaxy type shown by the symbols and colors.  Host type
  comes from \citet{gab09}, and is determined after subtracting the
  point-source AGN.  Spheroid hosts (with $n>2.5$) are shown by filled
  red circles, disk hosts ($n<2.5$) by blue bars, and irregular hosts
  ($A>0.5$) by green asterisks.  Error bars show the 1$\sigma$ errors
  in accretion rate and obscuration from the model fits.  Unobscured
  and rapidly accreting broad line AGNs prefer spheroid hosts, while
  weakly accreting AGNs are more frequently in disk galaxies and
  obscured AGNs are often disturbed.
\label{fig:unifiedhost}}
\end{figure}

Figure \ref{fig:unifiedhost} shows the host galaxy morphology with AGN
accretion rate and obscuration.  Each quadrant of the figure
corresponds to a different preferred host type.  All 5 of the
broad-line AGNs, which have the highest accretion rates among
unobscured ($N_H<10^{22}$~cm$^{-2}$) AGNs, have spheroid-dominated
hosts.  Weaker and obscured AGNs are frequently in disk galaxies, and
hosts of obscured ($N_H>10^{22}$~cm$^{-2}$) AGNs are more likely to be
disturbed.

The changes in host type with AGN accretion rate and obscuration are
consistent with merger-driven evolutionary models.  For example, while
weak AGNs might be fed by stochastic processes in disk galaxies
\citep{hh06}, mergers drive obscured AGNs \citep{san88}.  Further,
\citet{hop06} suggest that quasar feedback eventually blows out the
obscuring material, resulting in a post-merger spheroid with an
unobscured quasar.  However, we cannot rule out steady-state AGN
fueling: it may simply be that disturbed systems have more obscuring
material, and efficiently fueling an unobscured quasar requires a
massive spheroid.  In either case, our work is additional evidence for
a strong connection between galaxy properties and the growth and
obscuration of their supermassive black holes.

\acknowledgements The author acknowledges support from NASA HST grant
GO 12060.10-A, Chandra grant G08-9129A, and NSF grant AST- 0808133.
Helpful discussions with Knud Jahnke, Dale Kocevski, David Koo, and
Mike Koss contributed to the development of this work.  The COSMOS
team deserves enormous credit for producing the data used here.

\end{document}